\documentclass[preprint,superscriptaddress,preprintnumbers,showpacs]{revtex4}
\usepackage{amssymb}
\usepackage{graphicx}
\usepackage{dcolumn}
\usepackage{bm}

\begin{document}

\newcommand*{\sjtu}{INPAC, SKLPPC and Department of Physics, Shanghai Jiao Tong University,  Shanghai, China}\affiliation{\sjtu}
\newcommand*{\NTU}{CTS, CASTS and Department of Physics, National Taiwan University, Taipei, 106, Taiwan}\affiliation{\NTU}
\newcommand*{\NCTS}{Department of Physics, National Tsing Hua University, and National Center for Theoretical Sciences, Hsinchu, 300, Taiwan}\affiliation{\NCTS}

\title{The $\beta$ angle as the  CP violating phase in the CKM matrix}

\author{Guan-Nan Li}\affiliation{\sjtu}
\author{Hsiu-Hsien Lin}\affiliation{\NTU}
\author{Dong Xu}\affiliation{\sjtu}
\author{Xiao-Gang He}\email{hexg@phys.ntu.edu.tw}\affiliation{\sjtu}\affiliation{\NTU}\affiliation{\NCTS}

\begin{abstract}
The CKM matrix describing quark mixing with three generations can be
parameterized by three Euler mixing angles and one CP violating phase. In
most of the parameterizations, the CP violating phase chosen is not
a directly measurable quantity and is parametrization dependent. In
this work, we propose to use the most accurately measured CP violating
angle $\beta$  in the unitarity triangle
as the phase in the CKM matrix, and  construct an explicit 
$\beta$ parameterization.
We also derive an approximate Wolfenstein-like expression for this parameterization.

\end{abstract}


\maketitle

\noindent {\bf Introduction}

The mixing between different quarks is
described by an unitary matrix in the charged current
interaction of W-boson in the mass eigen-state of quarks,  the
Cabibbo~\cite{cabibbo}-Kobayashi-Maskawa~\cite{km}(CKM) matrix
$V_{\rm{CKM}}$, defined by
\begin{eqnarray}
L = -{g\over \sqrt{2}} \overline{U}_L \gamma^\mu V_{\rm CKM} D_L
W^+_\mu  + H.C.\;,
\end{eqnarray}
where $U_L = (u_L,c_L,t_L,...)^T$, $D_L = (d_L,s_L,b_L,...)^T$. For n-generations, $V =
V_{\rm CKM}$ is an $n\times n$ unitary matrix. With three generations, one can write
\begin{eqnarray}
V_{\rm CKM} = \left ( \begin{array}{lll}
V_{ud}&V_{us}&V_{ub}\\
V_{cd}&V_{cs}&V_{cb}\\
V_{td}&V_{ts}&V_{tb}
\end{array}
\right )\;.
\end{eqnarray}

A commonly used parametrization for mixing matrix with three generations of quark is given by~\cite{ck},
\begin{eqnarray}
V_{PDG} = \left(
\begin{array}{ccc}
c_{12}c_{13} & s_{12}c_{13} & s_{13}e^{-i\delta_{CK}}           \\
-s_{12}c_{23}-c_{12}s_{23}s_{13}e^{i\delta_{CK}} &
c_{12}c_{23}-s_{12}s_{23}s_{13}e^{i\delta_{CK}}  & s_{23}c_{13} \\
s_{12}s_{23}-c_{12}c_{23}s_{13}e^{i\delta_{CK}}  &
-c_{12}s_{23}-s_{12}c_{23}s_{13}e^{i\delta_{CK}} & c_{23}c_{13}
\end{array}
\right),\label{fp}
\end{eqnarray}
where $s_{ij}=\sin\theta_{ij}$ and $c_{ij}=\cos\theta_{ij}$ with $\theta_{ij}$ being angles rotating in flavor space
and $\delta_{CK}$ is the CP violating phase. We refer this as the CK parametrization. This form of parametrization was used
by Particle Data group as the standard parametrization\cite{yao}.

There are a lot of experimental data on the mixing pattern of
quarks. Fitting available data, the mixing angles and CP violating phase are determined to be~\cite{utfit}
\begin{eqnarray}
&&\theta_{12}=13.015^\circ\pm0.059^\circ,\quad
\theta_{23}=2.376^\circ \pm0.046^\circ,\quad
\theta_{13}=0.207^\circ\pm0.008^\circ, \nonumber\\
&&\delta_{CK}=69.7^\circ\pm3.1^\circ. \label{qangle}
\end{eqnarray}

The angles can be viewed as rotations in flavor spaces. But both the
angles and the phase in the CKM matrix are not directly measurable
quantities. There are different ways to parameterize the mixing
matrix. In different parametrizations, the angles and phase are
different. To illustrate this point let us study the original KM
parametrization~\cite{km},
\begin{eqnarray}
V_{KM} = \left ( \begin{array}{ccc} c_1& - s_1 c_3& -s_1 s_3\\s_1c_2&c_1c_2c_3 - s_2s_3 e^{i\delta_{KM}}&c_1c_2s_3 + s_2c_3 e^{i\delta_{KM}}\\
s_1s_2&c_1s_2c_3 + c_2 s_3 e^{i\delta_{KM}}& c_1s_2 s_3 - c_2c_3 e^{i\delta_{KM}}\end{array}
\right )\;.
\end{eqnarray}
Using the observed values for the mixing matrix, one would obtain
\begin{eqnarray}
&&\theta_1 = 13.016^\circ\pm0.003^\circ\;,\;\;\theta_2 = 2.229^\circ\pm0.066^\circ\;,\;\;\theta_3 =0.921^\circ\pm0.036^\circ\;,
\end{eqnarray}
and the central value of the CP violating phase angle is $\delta_{KM} = 88.2^\circ$.

\begin{figure}[htbp]
\centering
\includegraphics[width = 0.6\textwidth]{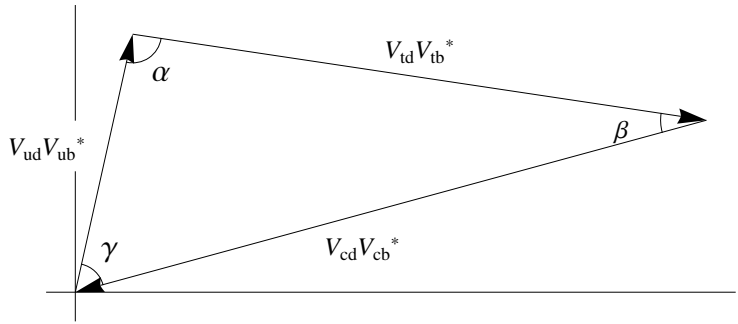}\hspace{0.5cm}
\caption{The unitarity triangle.}\label{fig}
\end{figure}

We see that the angles and phases in the CK and KM
parameterizations are indeed very different. The angles and phase
are parametrization dependent.
One can use this freedom to choose
a convenient parametrization to study. 
It is interesting to see whether all
quantities used to parameterize the mixing matrix can have well
defined physical meanings, that is, all are experimentally
measurable quantities, as have been done for several other
quantities related to mixing
matrices~\cite{jarlskog,triangle,fh,Pham:2011yy,qlc}. To this end we notice that
the magnitudes of the CKM matrix elements are already experimentally
measurable quantities, one can take them to parameterize the mixing
matrix. Experimentally there are also several
measurable angles which can signify CP violations. The famous ones
are the angles $\alpha$, $\beta$ and $\gamma$ in the unitarity
triangle defined by the unitarity condition
\begin{eqnarray}
V_{ud}V_{ub}^* + V_{cd}V_{cb}^* + V_{td}V^*_{tb}=0
\end{eqnarray}
In the complex plane, the above defines a triangle shown in Fig 1.
The unitarity of the CKM matrix actually defines six independent
triangle relations through: $\sum_j V_{ij}V^*_{kj} = 0$, and $\sum_j V_{ji}V^*_{jk} = 0$
for $i$ not equal to $k$. Among them, $i=d$ and $k=b$ case is the best studied experimentally
and the inner angles (phase angles) of the triangle independently measured.

The three inner angles defined by the triangle in Fig 1 can be expressed as
\begin{eqnarray}
\ {\alpha} = {\arg\left
(-{{V_{td}V_{tb}^*}\over{V_{ud}V_{ub}^*}}\right )}\;,\;\;
\ {\beta} = {\arg\left
(-{{V_{cd}V_{cb}^*}\over{V_{td}V_{tb}^*}}\right )}\;,\;\;
\ {\gamma} = {\arg\left
(-{{V_{ud}V_{ub}^*}\over{V_{cd}V_{cb}^*}}\right )}\;. \label{phases}
\end{eqnarray}

CP violation dictates that the area of this triangle to be non-zero. This
implies that none of the angles $\alpha$, $\beta$ and $\gamma$ can be zero.
Experimentally these three angles have been measured directly~\cite{yao},
$\alpha =(89.0^{+4.4}_{-4.2})^{\circ} $, $\beta =(21.1\pm0.9)^{\circ} $ and $\gamma =(73^{+22}_{-25})^{\circ} $.
These numbers are consistent with that obtained using the numerical numbers in eq. \ref{qangle}, $\alpha ={88.14}^{\circ} $, $\beta ={22.20}^{\circ} $ and $\gamma ={69.67}^{\circ} $.
Also the directly measured numbers are consistent with the
SM prediction $\alpha + \beta +\gamma = \pi$ in the CKM model with three generations.
Notice that the values $\alpha$, $\gamma$ are very close to the two phases
$\delta_{KM}$, $\delta_{CK}$, respectively. Among $\alpha$, $\beta$ and $\gamma$ angles, $\beta$ angle is the most accurately
measured one. It is therefore interesting to see if one can find a parameterization in which the CP violating phase is represented by the angle 
$\beta$. In the following we will discuss how one can obtain a parameterization using $\beta$ angle as the phase in the CKM matrix.

\noindent {\bf The $\beta$ angle parameterization}

Using eq.\ref{phases}, one can allocate the $\beta$ angle at different place, for example the following four ways  in which 
only one of the $V_{cd, cb, td, tb}$ relevant to the definition of $\beta$ is
complex and all others are real and positive,

\begin{eqnarray}
&&\beta_1)\;: \;\;(|V_{cd}|, |V_{cb}|, |V_{td}|, -|V_{tb}|e^{i\beta})\;,\nonumber\\
&&\beta_2)\;: \;\;(|V_{cd}|, |V_{cb}|, -|V_{td}|e^{-i\beta}, |V_{tb}|)\;,\nonumber\\
&&\beta_3)\;: \;\;(|V_{cd}|, -|V_{cb}|e^{-i\beta}, |V_{td}|, V_{tb}|)\;,\\
&&\beta_4)\;: \;\;(- |V_{cd}|e^{i\beta}, |V_{cb}|, |V_{td}|, |V_{tb}|)\;.\nonumber
\end{eqnarray}

The above defines four ways of parameterize the CKM matrix in which $\beta$ is explicitly the CP violating phase. 
These parameterizations are all equivalent. We will use $\beta_1$ for discussion. We have
\begin{equation}
V_{CKM}^{\beta_1} = \left(
\begin{array}{ccc}
|V_{ud}|& -{\frac{(|V_{ud}|^{2}-|V_{cb}|^{2})|V_{cd}|+|V_{cb}||V_{td}||V_{tb}|e^{i\beta}}{|V_{cs}||V_{ud}|}} &-{\frac{|V_{cb}||V_{cd}|-|V_{td}||V_{tb}|e^{i\beta}}{|V_{ud}|}}\\
|V_{cd}|&|V_{cs}| &|V_{cb}|\\
|V_{td}| & \frac{|V_{cb}||V_{tb}|e^{i\beta}-|V_{cd}||V_{td}|}{|V_{cs}|} & -|V_{tb}|e^{i\beta}\\
\end{array}
\right)\;,
\end{equation}\\
The CKM matrix is expressed explicitly in terms of modulus of matrix elements and the CP violating angle $\beta$.

 For this case, we can use $\beta$,~$|V_{cs}|$,~$|V_{cd}|$,~$|V_{td}|$ as independent variables, and express others as functions of them. We have
\begin{eqnarray*}
|V_{ud}|&=&\sqrt{1-|V_{cd}|^{2}-|V_{td}|^{2}},|V_{cb}|=\sqrt{1-|V_{cd}|^{2}-|V_{cs}|^{2}},\\
|V_{tb}|&=&\frac{|V_{cb}||V_{cd}||V_{td}|\cos{\beta}}{1-|V_{cd}|^{2}}+
\sqrt{(\frac{|V_{cb}||V_{cd}||V_{td}|\cos{\beta}}{1-|V_{cd}|^{2}})^{2}-\frac{|V_{cs}|^{2}(|V_{td}|^{2}-1)+|V_{cd}|^{2}|V_{td}|^{2}}{1-|V_{cd}|^{2}}}\;.
\end{eqnarray*}

The CP violating Jarlskog parameter $J$ \cite{jarlskog} is given by
\begin{eqnarray*}
J=|V_{cb}||V_{tb}||V_{cd}||V_{td}|\sin{\beta}.
\end{eqnarray*}

\noindent {\bf The $\beta$ and the Euler angle parameterizations}

Numerically, one finds that the approximate relations $\delta_{KM} \approx \alpha$ and $\delta_{CK}\approx \gamma$.
These can be understood easily by noticing the relations between them~\cite{fh,koide},
\begin{eqnarray}
&&\alpha
=\arctan({\sin \delta_{KM} \over x_{\alpha}-\cos\delta_{KM} }),\;\;\;\;x_{\alpha} = {c_1s_2s_3\over c_2c_3} = {|V_{ud}||V_{td}||V_{ub}|\over |V_{cd}||V_{us}|}=0.0006.\nonumber\\
&&\gamma=\arctan({\sin \delta_{CK} \over x_{\gamma}+\cos\delta_{CK} }),\;\;\;\;x_{\gamma}= {c_{12}s_{23}s_{13}\over s_{12} c_{23}} = {|V_{ud}||V_{cb}||V_{ub}|\over |V_{tb}||V_{us}|} =0.0006.\nonumber
\end{eqnarray}
Therefore, $\delta_{KM} + \alpha$ is approximately $\pi$, since $\alpha$ is close to $90^\circ$, $\delta_{KM}$ must also be close to $90^\circ$ and therefore $\delta_{KM} \approx \alpha$. It is also clear that $\delta_{CK}$ is approximately equal to $\gamma$.

One may wonder if there is a parameterization with three Euler angle and a phase where the phase is
close to $\beta$. We find indeed there are such prameterizations. An example is
provided by the parametrization $P4$ discussed in Ref.~\cite{xing}
where
\begin{equation}
 {V_{CKM}^{P4}}
 = \left(
\begin{array}{ccc}
c_{\theta}c_{\tau}& c_{\theta}s_{\sigma}s_{\tau}+ s_{\theta}c_{\sigma}e^{-i\varphi} & c_{\theta}c_{\sigma}s_{\tau}-s_{\theta}s_{\sigma}e^{-i\varphi}\\
-s_{\theta} c_{\tau}&-s_\theta s_\sigma s_\tau + c_\theta c_\sigma e^{-i\varphi}
 &-s_{\theta} c_{\sigma} s_\tau - c_\theta s_\sigma e^{-i\varphi}\\
 -s_{\tau} & s_{\sigma}c_{\tau} &
 c_{\sigma}c_{\tau}\\
\end{array}
\right).
\end{equation}
We have
\begin{eqnarray}
&&\beta
=\arctan({\sin \varphi \over x_{\beta}+ \cos\varphi }),\;\;\;\;x_{\beta} = {s_\theta c_\sigma s_\tau\over c_\theta s_\sigma} = {|V_{cd}||V_{tb}||V_{td}|\over |V_{ud}||V_{ts}|} = 0.0497.
\end{eqnarray}

\noindent {\bf A Wolfenstein-like Expansion}

It has proven to be convenient to use approximate formula such as the Wolfenstein parametrization\cite{wolfenstein}.
In the literatures different approximate forms have been proposed\cite{he-ma,ma-wolf}.
We now derive an approximate Wolfenstein-like parameterization in which $\beta$ is taken to the CP violating phase.

Setting $|V_{cd}|= \lambda$,
$|V_{td}| = b \lambda^3$, and  $|V_{cb}| = c \lambda^2$ with $\lambda=
0.2251\pm 0.0010$, and  $b=0.7685\pm0.0250 $,
$c= 0.8185\pm0.0176$. Rotating
the b-quark field by a phase $\pi -\beta$, we obtain to order
$\lambda^3$ for $V^{\beta_1}_{CKM}$
\begin{equation}
V_{CKM}^{\beta_{1}} \approx \left(
\begin{array}{ccc}
{1-{1\over 2}{\lambda}^2}& -\lambda & {\lambda}^3(c e^{-i\beta}-b)\\
\lambda & 1-{1\over 2}{\lambda}^2 & -c\lambda^2 e^{-i\beta}\\
b{\lambda}^3 & c\lambda^2{e}^{i\beta} & 1\\
\end{array}
\right).
\end{equation}

\noindent {\bf Conclusion}

To conclude, we have proposed a new parameterization 
using the most accuratly measured CP violating angle  $\beta$ in the unitarity triangle as the CP 
violating phase in the CKM matrix.  We find an Euler angle parameterization
in which the CP violating phase is very close to the angle $\beta$. 
We also derived a new Wolfenstein-like paramterization. Since $\beta$ is the most accurately measured among these three  angles in the unitarity triangle,
we therefore consider the $\beta$ parametrization the best one to use to provide information for CP violation.
\\

\noindent
{\bf Acknowledgement}
This work was supported in part by NSC and NCTS of ROC, NNSF(grant No:11175115) and Shanghai science and technology commission (grant no: 11DZ2260700) of PRC.

\end{document}